# A Chaotic Image Encryption Scheme Using Novel Geometric Block Permutation and Dynamic Substitution

Muhammad Ali[1], Jawad Ahmad[2] Muhammad Abdullah Hussain Khan[1], Safee Ullah[1], Mujeeb Ur Rehman[3], Syed Aziz Shah[4], and Muhammad Shahbaz Khan[5]

[1] Department of Electrical Engineering, HITEC University, Taxila, 47080, Pakistan
{20-ee-026, 20-ee-025}@student.hitecuni.edu.pk,
safee.ullah@hitecuni.edu.pk

[2] Cyber Security Center, Prince Mohammad Bin Fahd University, Al Khobar, 31952, Saudi Arabia
jahmad@pmu.edu.sa

[3] School of Computer Science and Informatics, De Montfort University, Leicester, LE1 9BH, UK
mujeeb.rehman@dmu.ac.uk

[4] Research Centre for Intelligent Healthcare, Coventry University, Coventry, CV1 5FB, UK
Syed.Shah@coventry.ac.uk

[5] School of Computing, Engineering and the Built Environment, Edinburgh Napier University, Edinburgh, EH10 5DT, UK
muhammadshahbaz.khan@napier.ac.uk

**Abstract.** In this digital era, ensuring the security of digital data during transmission and storage is crucial. Digital data, particularly image data, needs to be protected against unauthorized access. To address this, this paper presents a novel image encryption scheme based on a confusion diffusion architecture. The diffusion module introduces a novel geometric block permutation technique, which effectively scrambles the pixels based on geometric shape extraction of pixels. The image is converted into four blocks, and pixels are extracted from these blocks using L-shape, U-shape, square-shape, and inverted U-shape patterns for each block, respectively. This robust extraction and permutation effectively disrupts the correlation within the image. Furthermore, the confusion module utilises bit-XOR and dynamic substitution techniques. For the bit-XOR operation, 2D Henon map has been utilised to generate a chaotic seed matrix, which is bit-XORed with the scrambled image. The resultant image then undergoes the dynamic substitution process to complete confusion phase. A statistical security analysis demonstrates the superior security of the proposed scheme with high uncertainty and unpredictability, achieving an entropy of 7.9974 and a correlation coefficient of 0.0014. These results validate the proposed scheme's effectiveness in securing digital images.

## 1 Introduction

With the widespread use of digital technology and Internet-of-things (IoT) devices, the transmission of digital data, especially image data over unsecure channels has significantly increased. [1,2]. These images, which often contain sensitive personal, medical, or military information, are vulnerable to interception or modification during transmission and storage [3]. Therefore, securing digital images is essential to maintain the integrity of the information within them [4]. Image encryption plays a potential role in protecting digital images from unauthorized access and converts input image in a form that conceals the visual information in the image [5, 6]. Unlike text data, image data poses unique challenges, making traditional cryptographic methods like Data Encryption Standard (DES) [7], Triple DES (3DES) [8], and Advanced Encryption Standard (AES) [9] less effective due to latency and security concerns [10].

For an image encryption algorithm to be secure, it should have two important properties: diffusion and confusion [11]. Diffusion is the spread of the impact of one pixel of the plain image over various ciphered image pixels to hide the statistical properties of the plain image and is based on a secret key, while confusion alters pixel values, typically through substitution, complicating the relationship between the encrypted and original image [12]. Many methods rely solely on confusion, such as a single round single S-box method presented in [13]. But presence of pixel permutation with confusion strategies enhances the security of encryption algorithms. In addition, recent research trends have seen the integration of chaos theory into cryptography [14]. Chaotic systems are valuable for secure image communication due to their inherent randomness and unpredictability, characterized by non-repeating patterns, sensitivity to initial conditions [15, 16]. Hence, using chaotic maps to control the confusion and diffusion process adds more complexity and non-linearity to the encryption algorithms making them more unpredictable and unbreakable. In addition to chaos, there has been a recent interest in introducing quantum resistant cryptographic techniques to secure image data in post quantum era. Chaos theory is also being utilised in quantum cryptography algorithms to secure image data from future quantum attacks [17, 18].

Focusing on improving the encryption capabilities of the confusion-diffusion-based architectures, this paper introduces a new image encryption method that includes a novel geometric block permutation technique in conjunction with a strong confusion module comprising of dynamic substitution. The main contributions of the proposed scheme are:

1. A novel geometric block permutation scheme is proposed that scrambles the pixels based on different geometric shapes to extract pixels from the original image and rearrange them randomly. This effectively breaks the inherent correlation between the pixels and makes the proposed encryption scheme more secure.

2. A robust chaotic confusion module is applied after the proposed permutation scheme. The confusion module consists of two steps; (a) a chaotic seed matrix is generated using the Hénon Map and it is bit-XORed with the permuted image, (b) a dynamic substitution process is applied on the bit-XORed image to produce a secure encrypted image.

## 2 The Proposed Encryption Scheme

The proposed encryption algorithm begins with the chaotic key generation module, utilizing the Hénon map for creating a seed matrix used in the bit-XOR operations, and a chaotic sequence for the selection of S-boxes during dynamic substitution process. Furthermore, the proposed algorithm encompasses diffusion and confusion phases to ensure secure encryption. The proposed encryption algorithm is given in Figure 1 and the detailed steps are described as follows:

### 2.1 Step 1: Generation of Chaotic Sequences Utilizing Hyper Chaotic Map

**Generation of Chaotic Seed Matrix:** The 2D Hénon map is utilized to generate random matrices for Bit-wise XOR operations and S-Box selection operation. The equations of 2-D Hénon map are:

$$x_{n+1} = 1 - ax_n^2 + y_n \tag{1}$$

$$y_{n+1} = bx_n \tag{2}$$

Where a and b the control parameters.

To generate the chaotic seed matrix R1 with dimensions $(M \times N)$ such as $(256 \times 256)$, the map is iterated to produce numbers with decimal points or floating-point numbers between 0 and 1. To eliminate fractions, each number is scaled and rounded to get integers with finite precision sequence, which is denote as $R_1'$.

$$R_1' = round(R_1 \times 10^5 \ ) \tag{3}$$

However, to ensure that the values of resultant matrix are within the range of 0 to 255, a modulus operation is applied. The modulus operation limits the values within the desired range, resulting in random matrix of $M\times$ N dimensions.

$$R_1' = mod(R_1', 255) \tag{4}$$

By following these steps, a random array is generated, ensuring the values are within the specified range and are suitable for bit-wise XOR operations.

**Generation of S-Box Selection Sequence:** For S-Box selection, a sequence having dimensions equal to the length of the image is generated through the 2D Hénon map.In a similar manner as described earlier, the map is iterated and a random sequence is generate, which has fractional values between 0 and 1. To eliminate fractions, each number is scaled and rounded to get integers with finite precision. The obtained sequence is denoted as $S_1'$.

$$S_1' = round\left(R_1 \times 10^5\right) \tag{5}$$

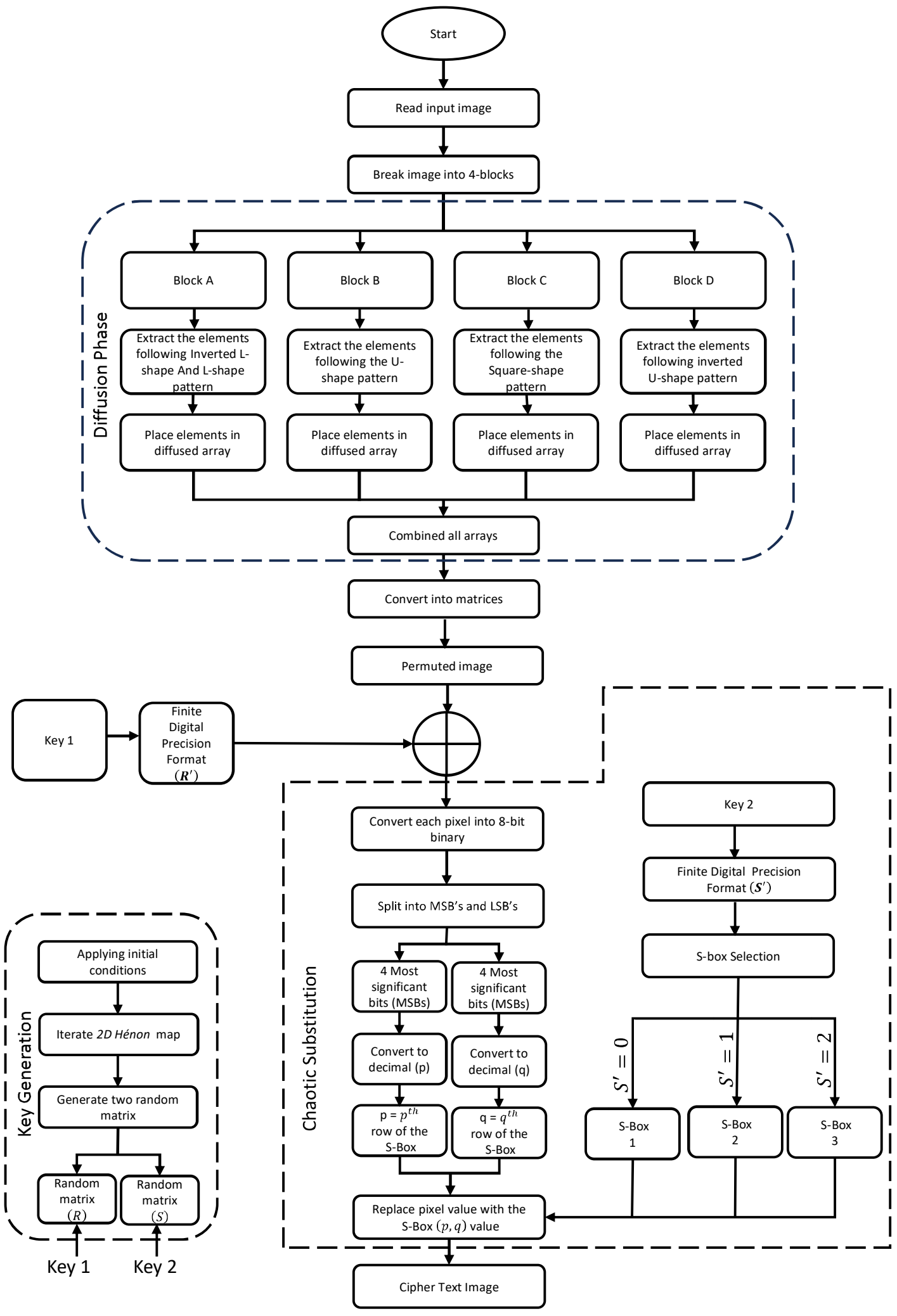

Fig. 1: The proposed image encryption scheme

However, to ensure that the S-Box selection matrix contain values within the range of 1 to 3, a modulus operation is applied. This operation helps us limit the values within the desired range, resulting in random array.

$$S_1' = mod\,(R_1', 3) \tag{6}$$

Through these steps, a random sequence is generated, ensuring that it contains values within the specified range and is suitable for S-Box selection operation.

### 2.2 Step 2: Geometric Block Permutation Technique

The process begins by decomposing the input image of size $256 \times 256$ into four equal-sized blocks named as Block A, Block B, Block C, and Block D. Then flip each block by 180° as shown in Figure 2. From each block, specific shapes are extracted and stored in a new array using the following sequence. The visual interpretation of the geometric block permutation scheme is given in 3.

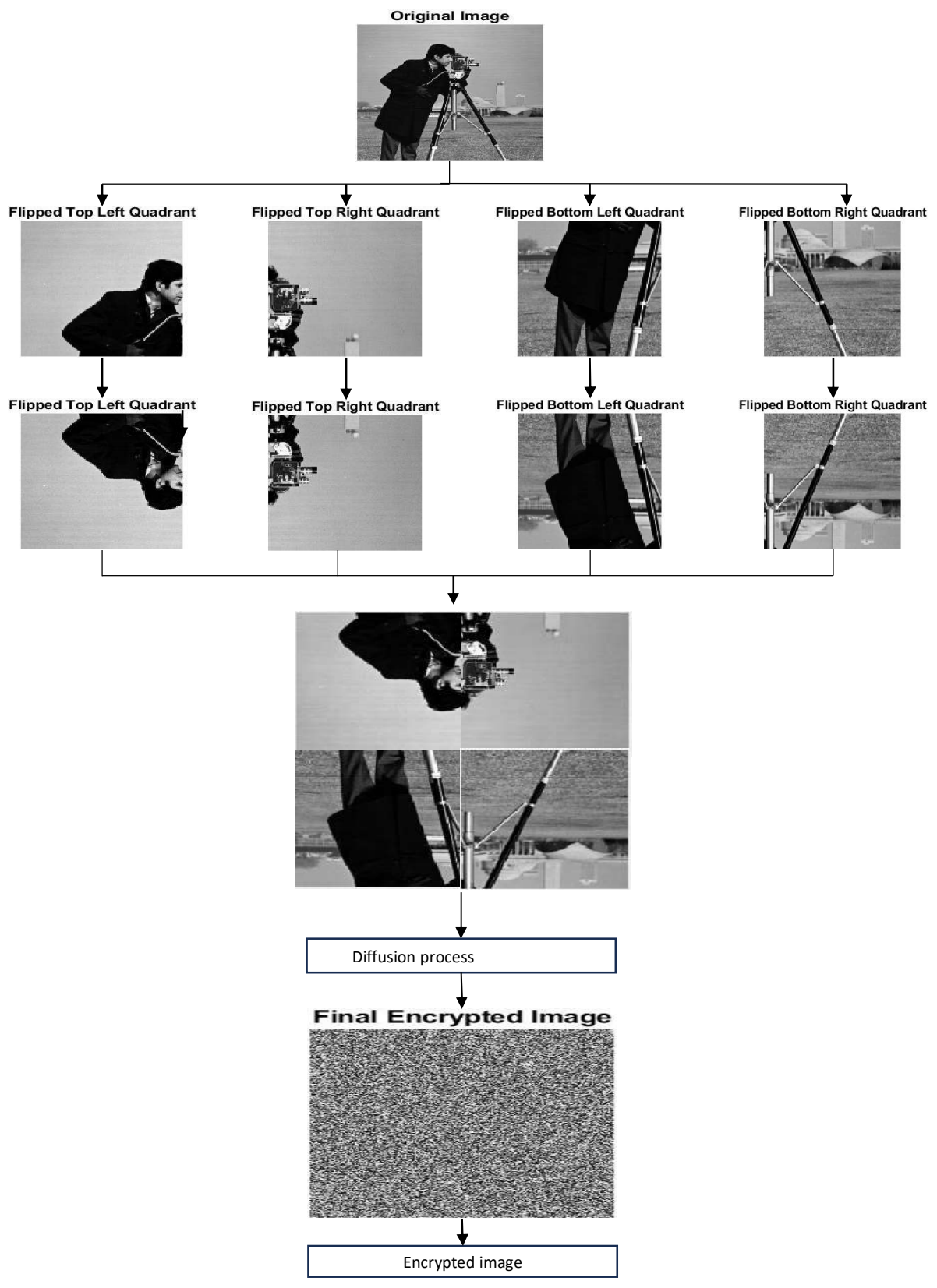

Fig. 2: Decomposing Original Image and Diffused Image

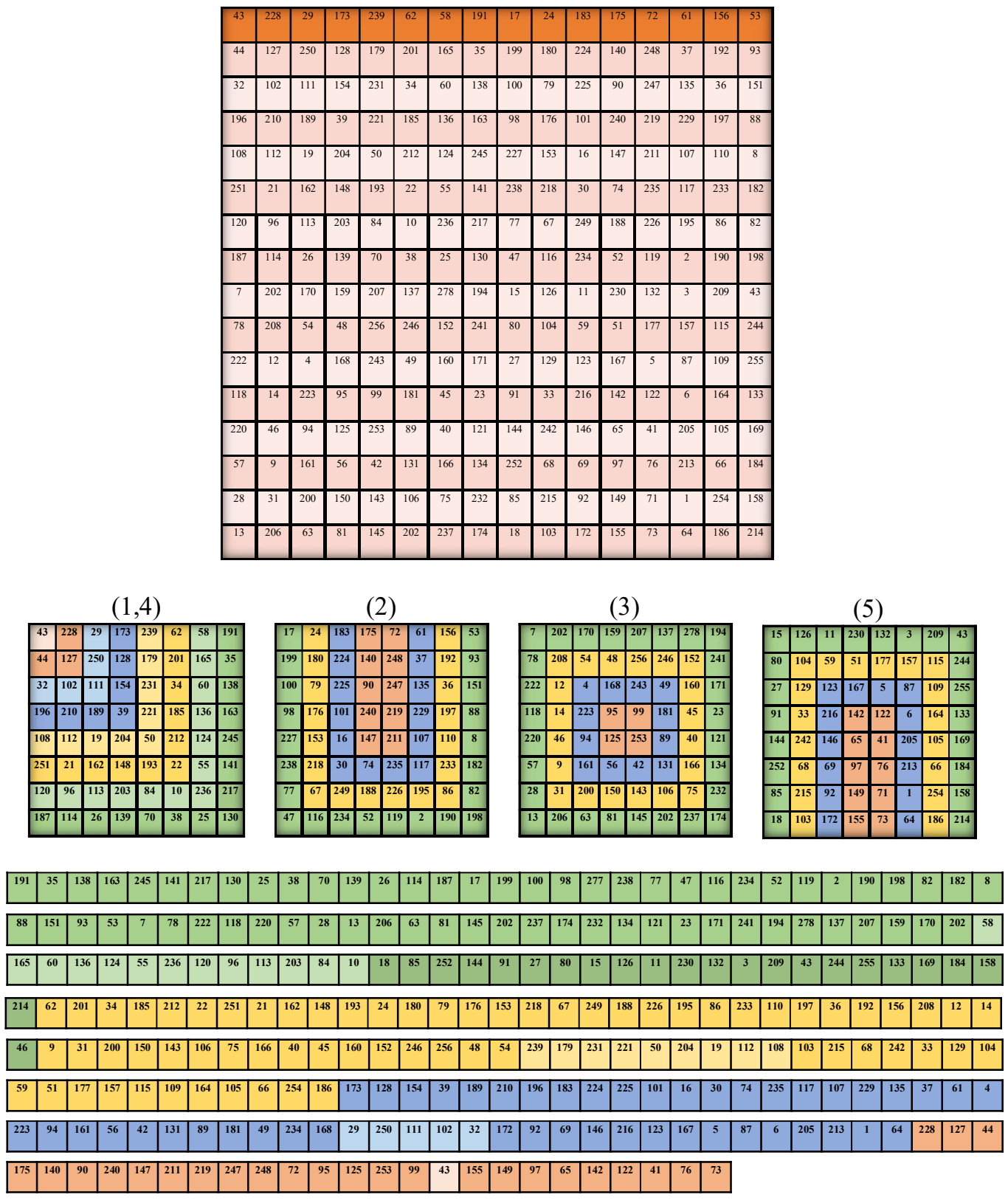

Fig. 3: The proposed geometric block permutation

- **Inverted L-Shape from Block A:** The L-shape extraction starts with extracting pixels from the $(N/2)$ column and continue extracting pixels from the end of $(M/2)$ row. With each iteration, the column and row indices are reduced by one$[(N/2)-1,(M/2)-1]$. This pattern continues until the entire Block A is covered. The extracted elements are then placed into a diffused array.
- **U-Shape from Block B:** It starts with extracting pixels from the first column, then moving towards y$(M/2)$ row. After completing the row, extraction of pixels continues from the end of the $(N/2)$ column until the column is fully covered. With each iteration, an increment of one is added in first column index and both the first and last column and row indices are reduced by one $[1+1,(N/2)-1,(M/2)-1]$, continuing this pattern until the entire Block B is processed. The extracted elements are then placed into the diffused array, next to the pixels from the previously extracted shapes.

- **Square Shape from Block C:** It starts with extracting pixels from the first and last columns$[1, (N/2)]$, as well as the first and last rows$[1, (M/2)]$, from Block C. With each iteration, the last column and last row indices are reduced by one $[(N/2) - 1, (M/2) - 1]$, while incrementing the first column and first row indices by one $[1 + 1, 1 + 1]$. This pattern continues until the entire Block C is processed. The elements that are extracted are placed into the diffused array, next to the pixels from the previously extracted shape.
- **L-Shape from Block A:** This step begins by extracting the pixels from the first column and then proceeds to extract from $(M/2)$ row. With each iteration, the first column and first row indices are incremented by one. Continue following this pattern until the entire Block A is covered. The extracted elements are then placed into the diffused array, next to the pixels from the previously extracted shapes.
- **Inverted U-Shape from Block B:** This step starts by extracting pixels from end of the first column, then move to first row. After completing the row, pixels from the start of the $(N/2)$ column are extracted until the column is fully covered. With each iteration, an increment of one is added in the first column index and both the first column and row indices are reduced by one $[1+1, (N/2)-1, (M/2)-1]$, continuing this pattern until the entire Block B is processed. The extracted elements are then placed into the diffused array, next to the pixels from the previously extracted shapes.

By tracing the patterns of shape extraction and placement in the diffused array of size $1 : M \times N$, every pixel within image blocks is scrambled efficiently, ensuring effective permutation. Once all pixels are placed in the array, it is then reshaped back into a matrix, resulting in the successfully diffused image, ready for subsequent encryption steps.

### 2.3 Step 3: Bit-XOR Operation

Furthermore, a bit-wise XOR operation is employed to the scrambled image utilizing a chaotic seed matrix generated by the 2D Hénon map. Each pixel of the permuted image is bit-XORed with the each corresponding value of the chaotic seed matrix. This operation induces confusion in the permuted image. The change in pixel values by applying the XOR operation enhances the overall security of the encryption process.

### 2.4 Step 4: Dynamic Substitution

A dynamic substitution technique is applied on the bit-XORed image adding a double layer of confusion. This technique utilises three different S-boxes, i.e., S-box1 (AES S-box), S-box2 (Hussain's S-box), and S-box3 (Gray S-box). An S-box selection matrix is generated using the Hénon map, that determines which S-box will be selected to replace the pixel value. The substitution process starts with reading the bit-XORed image and converting the pixel value into an 8-bit binary format, which includes: 4-bit Most Significant Bits (MSBs) and 4-bit Least Significant Bits (LSBs). The rows and columns of the selected S-box are represented by MSBs and LSBs. Then the respective value from the S-box selection matrix is read.If 0 appears in the S-Box selection matrix, the value is picked from S-box1 using the MSBs and LSBs and this value is substituted into the bit-XORed matrix. If the value is 1 or 2, the corresponding S-box (S-box2 or S-box3) is chosen to pick the value to be substituted from.

## 3 Results and Statistical Security Analysis

This section presents results of the proposed encryption scheme with its statistical security analysis performed on the Cameraman test image. The method effectively conceals image data, as shown by low correlation and high entropy values. The analysis confirms that the scheme enhances security by improving data scrambling and hence, improves its protection against unauthorized access.

### 3.1 Histograms

The histogram represents the distribution of pixel values at different grey levels within an image. This information is crucial as it can lead back to the original image. An encrypted image should have equalised histogram or a histogram having uniform distribution of pixel values. The uniform distribution hides the information of shapes/patterns in the original image. It can be seen in Figure 4 that the histogram of the cipher images have equally distributed pixel intensities and uniform distribution validating the security of the proposed scheme.

### 3.2 Information Entropy

Entropy measures the amount of randomness or uncertainty in encrypted images. Higher entropy indicates high randomness of information and enhances security level. Entropy analysis helps in assessing the encryption strength by evaluating the unpredictability in the image. For the proposed scheme, the entropy of the encrypted image is calculated as 7.9969, which indicates a higher uncertainty and strong unpredictability. The result of entropy is given in Table 1 and it is calculated as:

The entropy of an encrypted image $E_{img}$ is calculated using the following formula:

$$E_{img} = -\sum_{p_i \in P} f(p_i) \cdot \log_2(f(p_i)) \tag{7}$$

Where:

- $E_{img}$ is the entropy of the image.
- $P$ is the set of all unique pixel intensity levels or color values in the image.
- $p_i$ is an individual pixel intensity or color value.
- $f(p_i)$ is the probability of occurrence of pixel value $p_i$, calculated as $\frac{N(p_i)}{N_{total}}$, where $N(p_i)$ is the number of occurrences of $p_i$ and $N_{total}$ is the total number of pixels in the image.
- $\log_2$ represents the logarithm to the base 2.

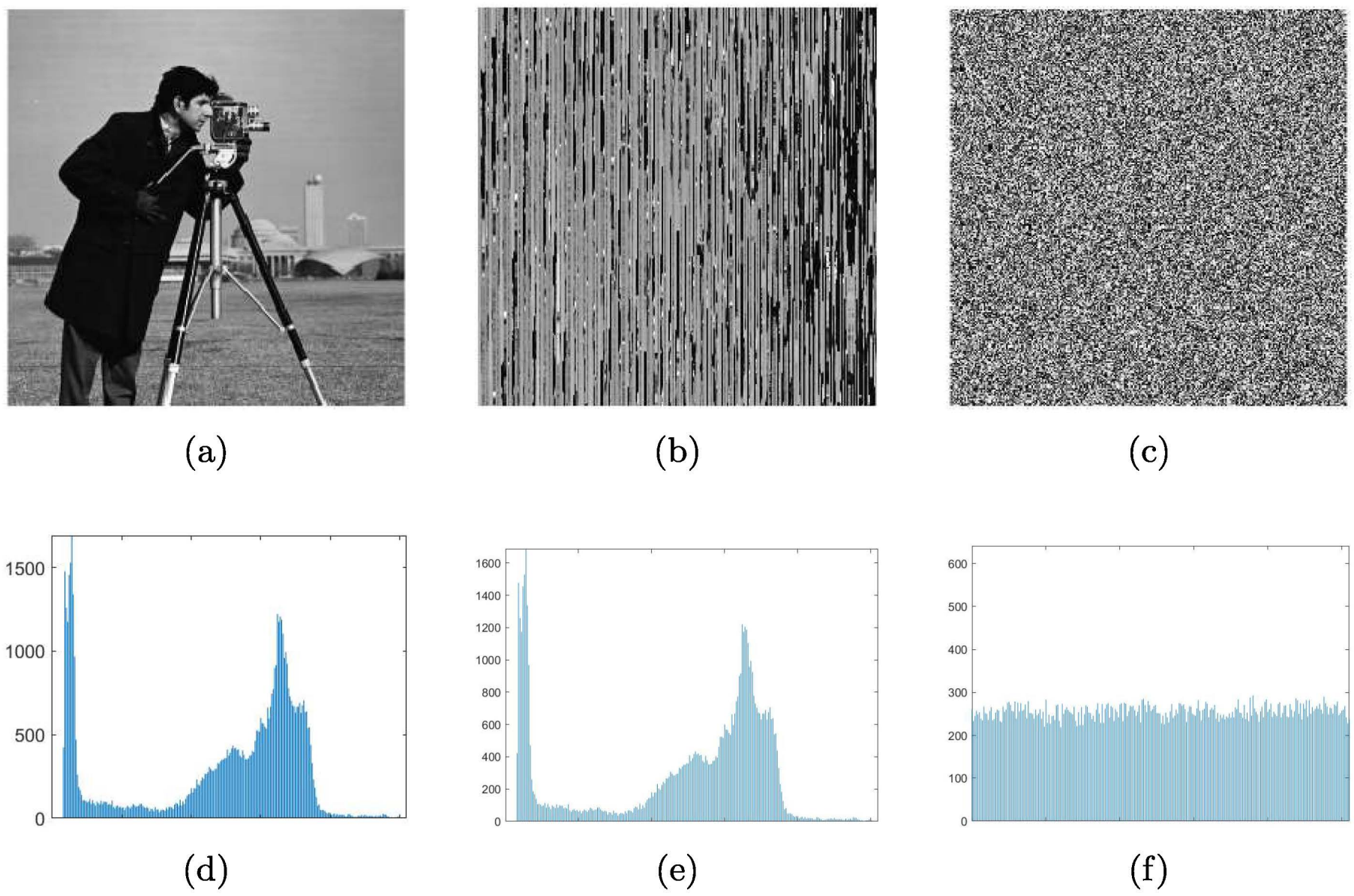

Fig. 4: Histogram Analysis of the proposed scheme, (a and d). Original Image and its histogram, (b and e). Diffused Image and its histogram, (c and f). Encrypted image and its Histogram.

### 3.3 Homogeneity

Homogeneity in image encryption refers to the uniformity of pixel values across the encrypted image. Higher homogeneity indicates a more consistent distribution of pixel intensities, potentially reducing the visual distortion introduced by encryption. It ensures that encrypted images maintain coherence and are recognizable.The homogeneity of an encrypted image $H_{img}$ is calculated using the following formula and the result of homogeneity analysis is given in Table 1.

$$H_{img} = \sum_{i=1}^{M}\sum_{j=1}^{N} \frac{1}{1 + |I(i,j) - I(i',j')|} \tag{8}$$

Where:

- $H_{img}$ is the homogeneity of the image.
- $M$ and $N$ represent the dimensions of the image (rows and columns).
- $I(i,j)$ is the intensity or color value of the pixel at position $(i,j)$.
- $I(i',j')$ is the intensity or color value of a neighboring pixel.
- $|I(i,j) - I(i',j')|$ is the absolute difference in intensity between the pixel and its neighboring pixel.

### 3.4 Energy

Energy in encrypted images gauges pixel intensity fluctuations, indicating texture. Lower values suggest smoother pixel distribution often indicative of robust encryption.

The energy of an encrypted image $E_{energy}$ is calculated using the following formula:

$$E_{energy} = \sum_{i=1}^{M} \sum_{j=1}^{N} I(i, j)^2 \tag{9}$$

Where:

- $E_{energy}$ is the energy of the image.
- $M$ and $N$ represent the dimensions of the image (rows and columns).
- $I(i, j)$ is the intensity or color value of the pixel at position $(i, j)$.

### 3.5 Contrast

Contrast in image encryption refers to the difference in pixel intensity values across the image, indicating variation in brightness. Higher contrast implies a wider range of intensity values, potentially aiding in image clarity or detail. Assessing contrast in encrypted images helps gauge the preservation of visual information during encryption processes. Maintaining appropriate contrast levels is crucial for ensuring the perceptual quality and effectiveness of encrypted images. The contrast of an encrypted image $C_{img}$ is calculated using the following formula:

$$C_{img} = \sum_{i=1}^{M} \sum_{j=1}^{N} \left(I(i, j) - I(i', j')\right)^2 \tag{10}$$

Where:

- $C_{img}$ is the contrast of the image.
- $M$ and $N$ represent the dimensions of the image (rows and columns).
- $I(i, j)$ is the intensity or color value of the pixel at position $(i, j)$.
- $I(i', j')$ is the intensity or color value of a neighboring pixel.
- $\left(I(i, j) - I(i', j')\right)^2$ is the squared difference between the pixel and its neighboring pixel.

### 3.6 Correlation Coefficients

Correlation in image encryption quantifies the linear relationship between pixel intensities in an encrypted image. Higher correlation values indicate stronger linear dependence, potentially preserving image structures. On the other hand, lower correlation values (ideally close to zero), represent no relation between the neighbouring pixels. A securely encrypted image should have zero or near to zero correlation value. The result of the correlation value is given in Table 1. It is evident that the correlation value is very close to zero.

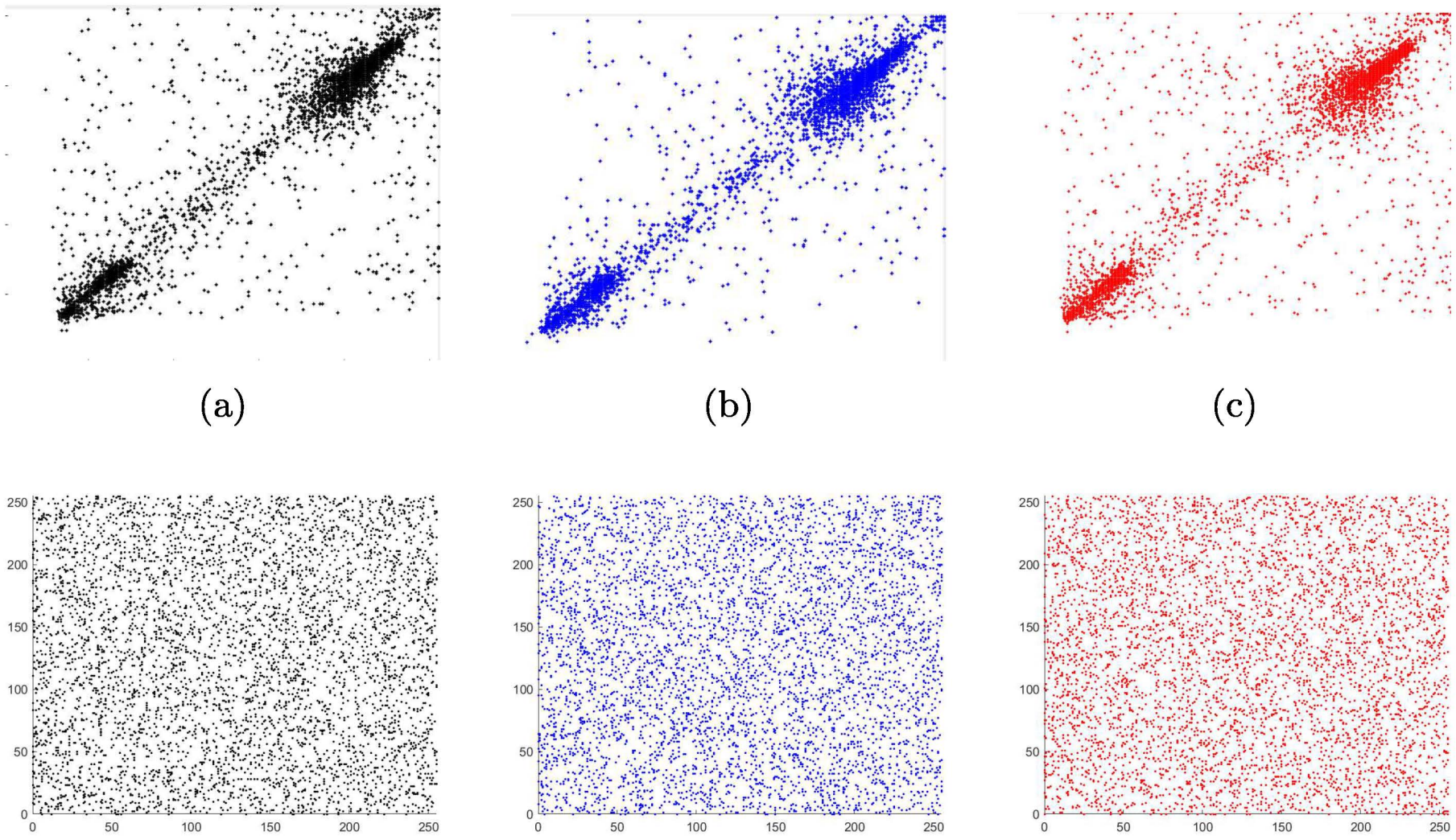

(a) (b) (c)

(d) (e) (f)

Fig. 5: Correlation Analysis of the proposed scheme.

Table 1: Results of the security analysis

| | | |
|---|---|---|
| **Entropy Analysis** | Entropy | 7.9974 |
| **GLCM** | Correlation | 0.0086 |
| | Contrast | 10.5089 |
| | Energy | 0.0156 |
| | Homogeneity | 0.3902 |

## 4 Conclusion

This paper presented a novel and secure image encryption algorithm having a confusion-diffusion architecture. The diffusion phase introduced a novel geometric block permutation technique. The proposed permutation technique efficiently scramble the pixels of the input images by dividing the image in 4 blocks and then applying different shapes based pixel extraction technique to effectively permute the pixels. The confusion, on the other hand, architecture consisted of bit-XOR and dynamic substitution techniques. The proposed scheme performed exceptionally well in the statistical security analysis and showcased its efficacy in encrypting images effectively.